\documentclass[]{spie}  

\usepackage{amsmath,amsfonts,amssymb}
\usepackage{graphicx}
\usepackage[colorlinks=true, allcolors=blue]{hyperref}
\usepackage{adjustbox}
\usepackage{caption}
\usepackage{subcaption}
\usepackage{multirow}
\usepackage{upgreek}
\usepackage[dvipsnames]{xcolor}

\newcommand\phip{$\phi_{pair}$}
\newcommand\epsp{$\epsilon_{pair}$}
\newcommand\EBL{E$\rightarrow$B Leakage}

\title{Polarization Calibration of the BICEP3 CMB polarimeter at the South Pole}

\author[a]{J.~Cornelison}
\affil[a]{Center for Astrophysics $\vert$ Harvard \& Smithsonian, Cambridge, MA 02138, U.S.A}

\author[b]{P.~A.~R.~Ade}
\affil[b]{School of Physics and Astronomy, Cardiff University, Cardiff, CF24 3AA, United Kingdom}

\author[c,d]{Z.~Ahmed}
\affil[c]{SLAC National Accelerator Laboratory, 2575 Sand Hill Road, Menlo Park, CA 94025}
\affil[d]{Kavli Institute for Particle Astrophysics and Cosmology, Stanford University, 452 Lomita Mall, Stanford, CA 94305}

\author[e]{M.~Amiri}
\affil[e]{Department of Physics and Astronomy, University of British Columbia, Vancouver, British Columbia, V6T 1Z1, Canada}

\author[a,f]{D.~Barkats}
\affil[f]{Institut Laue-Langevin, 38042 Grenoble Cedex 9, France}

\author[g]{R.~Basu Thakur}
\affil[g]{Department of Physics, California Institute of Technology, Pasadena, California 91125, USA}

\author[h]{C.~A.~Bischoff}
\affil[h]{Department of Physics, University of Cincinnati, Cincinnati, Ohio 45221, USA}

\author[g,i]{J.~J.~Bock}
\affil[i]{Jet Propulsion Laboratory, Pasadena, California 91109, USA}

\author[a]{H.~Boenish}
\author[j]{E.~Bullock}
\affil[j]{Minnesota Institute for Astrophysics, University of Minnesota, Minneapolis, 55455, USA}

\author[k]{V.~Buza}
\affil[k]{Kavli Institute for Cosmological Physics, University of Chicago, Chicago, IL 60637, USA}

\author[l]{J.~R.~Cheshire}
\affil[l]{School of Physics and Astronomy, University of Minnesota, Minneapolis, 55455, USA}

\author[m]{J.~Connors}
\affil[m]{National Institute of Standards and Technology, Boulder, Colorado 80305, USA}

\author[l]{M.~Crumrine}
\author[d,c,n]{A.~Cukierman}
\affil[n]{Department of Physics, Stanford University, Stanford, California 94305, USA}

\author[m]{E.~Denison}
\author[a]{M.~Dierickx}
\author[o]{L.~Duband}
\affil[o]{Service des Basses Temp\'{e}ratures, Commissariat \`{a} lEnergie Atomique, 38054 Grenoble, France}

\author[a]{M.~Eiben}
\author[e]{S.~Fatigoni}
\author[p,q]{J.~P.~Filippini}
\affil[p]{Department of Physics, University of Illinois at Urbana-Champaign, Urbana, Illinois 61801}
\affil[q]{Department of Astronomy, University of Illinois at Urbana-Champaign, Urbana, Illinois 61801, USA}

\author[l]{S.~Fliescher}
\author[d,n]{N.~Goeckner-Wald}
\author[a]{D.~C.~Goldfinger}
\author[n]{J.~A.~Grayson}
\author[a]{P.~Grimes}
\author[l]{G.~Hall}
\author[e]{M.~Halpern}
\author[a]{S.~A.~Harrison}
\author[c,d]{S.~Henderson}
\author[g,i]{S.~R.~Hildebrandt}
\author[m]{G.~C.~Hilton}
\author[m]{J.~Hubmayr}
\author[g]{H.~Hui}
\author[c,d,n,m]{K.~D.~Irwin}
\author[g,n]{J.~Kang}
\author[k]{K.~S.~Karkare}
\author[n]{E.~Karpel}
\author[g]{S.~Kefeli}
\author[n]{S.~A.~Kernasovskiy}
\author[a,r]{J.~M.~Kovac}
\affil[r]{Department of Physics, Harvard University, Cambridge, MA 02138, USA}

\author[n,c,d]{C.~L.~Kuo}
\author[l]{K.~Lau}
\author[k]{E.~M.~Leitch}
\author[i]{K.~G.~Megerian}
\author[g]{L.~Minutolo}
\author[g]{L.~Moncelsi}
\author[l]{Y.~Nakato}
\author[s]{T.~Namikawa}
\affil[s]{Department of Applied Mathematics and Theoretical Physics, University of Cambridge, Cambridge CB3 0WA, UK}

\author[g,i]{H.~T.~Nguyen}
\author[g,i]{R.~O'Brient}
\author[n]{R.~W.~Ogburn~IV}
\author[h]{S.~Palladino}
\author[l]{N.~Precup}
\author[o]{T.~Prouve}
\author[k,l]{C.~Pryke}
\author[a]{B.~Racine}
\author[m]{C.~D.~Reintsema}
\author[a]{S.~Richter}
\author[g]{A.~Schillaci}
\author[a]{B.~L.~Schmitt}
\author[l]{R.~Schwarz}
\author[j]{C.~D.~Sheehy}
\author[g]{A.~Soliman}
\author[a,q]{T.~St.~Germaine}

\author[g]{B.~Steinbach}
\author[b]{R.~V.~Sudiwala}
\author[t]{G.~P.~Teply}
\affil[t]{Department of Physics, University of California at San Diego, La Jolla, California 92093, USA}
\author[d,n]{K.~L.~Thompson}
\author[n]{J.~E.~Tolan}
\author[b]{C.~Tucker}
\author[i]{A.~D.~Turner}
\author[p,q]{C.~Umilt\`{a}}
\author[k]{A.~G.~Vieregg}
\author[g]{A.~Wandui}
\author[i]{A.~C.~Weber}
\author[e]{D.~V.~Wiebe}
\author[l]{J.~Willmert}
\author[a,r]{C.~L.~Wong}
\author[c,d,n]{W.~L.~K.~Wu}
\author[n]{E.~Yang}
\author[n,c,d]{K.~W.~Yoon}
\author[d,c,n]{E.~Young}
\author[n]{C.~Yu}
\author[a]{L.~Zeng}
\author[g]{C.~Zhang}
\author[g]{S.~Zhang}

\begin{document} 
\maketitle
\newpage
\begin{abstract}
The BICEP3 CMB Polarimeter is a small-aperture refracting telescope located at the South Pole and is specifically designed to search for the possible signature of inflationary gravitational waves in the Cosmic Microwave Background (CMB). 
The experiment measures polarization on the sky by differencing the signal of co-located, orthogonally polarized antennas coupled to Transition Edge Sensor (TES) detectors.
We present precise measurements of the absolute polarization response angles and polarization efficiencies for nearly all of BICEP3’s $\sim800$ functioning polarization-sensitive detector pairs from calibration data taken in January 2018. Using a Rotating Polarized Source (RPS), we mapped polarization response for each detector over a full 360 degrees of source rotation and at multiple telescope boresight rotations from which per-pair polarization properties were estimated.
In future work, these results will be used to constrain signals predicted by exotic physical models such as Cosmic Birefringence.
\end{abstract}


\section{INTRODUCTION}
\label{sec:intro}
Conservation of parity is part of the foundation of all modern physics.
However, it has been found that conservation of parity is violated in some interactions that involve the weak force \cite{1956Lee,1957Wu}. 
Unification of the electromagnetic and weak forces at high energies suggests that the electromagnetic force may violate parity conservation as well.
Additions to the theoretical framework of modern physics have been investigated in order to hypothesize the characteristic observables in scenarios where parity is not conserved in electromagnetism.
The Chern-Simons Lagrangian is one such example in which parity symmetry is violated and physically manifests in a divergence in the speed of circularly polarized light depending on polarization direction\cite{1990carroll}.
Since linear polarization can be described as the superposition of two circularly polarized modes of opposite handedness, the lagging of one mode compared to the other as they propagate through space manifests as a rotation by some angle, $\alpha$, on the axis of the linear polarization.
As such, a birefringent signal would be largest at the farthest observable distances in the universe, namely the Cosmic Microwave Background (CMB).

The CMB is a mostly uniform $2.725\,\text{K}$ blackbody radiator with anisotropies in temperature of $\mathcal{O}(30\,\upmu\text{K})$\cite{1992smoot} and with even fainter curl-less E-mode polarization anisotropies of $\mathcal{O}(1\,\upmu\text{K})$\cite{2002kovac}.
Cosmic birefringence in the universe would have the effect of creating \EBL{} in which primordial CMB E-modes are rotated into B-mode polarization and vice versa.
In power spectral space, this shows up as power in EB and TB cross-correlations which are canonically zero in the $\Lambda$CDM model.
For CMB experiments in which the primary focus is detecting primordial B-modes, \EBL{} of this nature is removed through an EB/TB minimization procedure\cite{2009wu} often referred to as ``self-calibration"\cite{2013selfcal}.
A miscalibration of the telescope's own polarization axis (often characterized as $\Delta \psi$) also produces an apparent rotation in linear polarization and introduces the same signature in EB and TB spectra\cite{2013selfcal}.
Thus, the challenge of constraining birefringence for CMB experiments is breaking the degeneracy between EB and TB power caused by instrumental effects and legitimate birefringent signals which requires both precise and accurate measurements of the absolute polarization orientation of the instrument.
Throughout the years, many experiments have reported constraints on birefringent signals, the strongest constraints of which come from ACTPol (see Table \ref{tab:bcconstraints}).

\begin{table}[h]

    \centering
\begin{adjustbox}{width={\textwidth},totalheight={\textheight},keepaspectratio}
\begin{tabular}{ ccccc }

\hline
\hline
\textbf{Experiment} & \textbf{Frequency (GHz)} & \textbf{$\ell$ range} & \textbf{$\alpha (^\circ)\pm\text{(stat)}\pm\text{(sys)}$} & \textbf{Calibration Method}\\
\hline

\hline
\multirow{2}{*}{QUaD\cite{2009wu}} & 100 & \multirow{2}{*}{200-2000} & $-1.89\pm2.24 (\pm0.5)$ & \multirow{2}{*}{polarized source}\\
& 150 && $+0.83\pm0.94 (\pm0.5)$ &\\

\hline
BOOM03\cite{2009pagano} & 143 & 150-1000 & $-4.3\pm4.1$ & pre-flight polarized source \\

\hline
ACTPol\cite{2009finelli} & 146 & 500-2000 & $-0.2\pm0.5(-1.2)$ & ''As-Designed'' \\

\hline
WMAP7\cite{2011Komatsu} & 41+61+94 & 2-800 & $-1.1\pm1.4 (\pm1.5)$ & pre-launch polarized source / Tau A \\

\hline
BICEP2\cite{2013aiken} & 150 & 30-300 & $-1\pm0.2(\pm1.5) $ & Dielectric Sheet \\

\hline
\multirow{3}{*}{BICEP1\cite{2014selfcal}} & \multirow{3}{*}{100+150} & \multirow{3}{*}{30-300} & $-2.77\pm0.86(\pm1.3)$ & Dielectric Sheet\\ 
&&& $-1.71\pm0.86(\pm1.3)$ & Polarized Source \\
&&& $-1.08\pm0.86(\pm1.3)$ & ''As-designed'' \\

\hline
POLARBEAR\cite{2014polarbear} & 150 & 500-2100 & $-1.08\pm0.2 (\pm0.5)$ & Tau A\\

\hline
\textit{Planck}\cite{2016planckparity} & 30-353 & 100-1500 & $0.35\pm0.05 (\pm0.28)$ & pre-flight source / Tau A$^\dagger$\\
\hline
ACTPol\cite{2020choi} & 150 & 600-1800 & $-0.07\pm0.09$ & metrology+modeling+planet obs. \\

\hline
\hline
\end{tabular}
\end{adjustbox}
\caption{Uniform cosmic birefringence constraints from CMB experiments ordered chronologically by publication.}
{\footnotesize $^\dagger$ Calibration of polarization orientations were completed pre-flight in the near field and confirmed with celestial source Tau A within uncertainties \cite{2010planckgroundcal,2016planckspacecal}}.
\label{tab:bcconstraints}
\end{table}

The BICEP3 polarimeter is one of a family of experiments conducted by the BICEP Collaboration at the South Pole.
It is a $0.5\,\text{m}$ refracting telescope capable of precisely measuring the fluctuations in polarization of the Cosmic Microwave Background at a frequency of $95\,\text{GHz}$ \cite{2014B3Zeesh}.
BICEP3 is capable of fielding up to 1200 optically active detector pairs each consisting of two co-located orthogonally polarized detectors such that, when combined, each on-sky detector pair fully measures Stokes $Q$ and $U$.
The detector pairs are grouped into 8-by-8 square tiles which can be removed and replaced without interfering with other tiles on the focal plane unit (FPU) (see Fig. 4 of Hui et al. (2016)\cite{2016B3Hui}).
The telescope itself is mounted on a three-axis mount that moves in Azimuth and Elevation with an additional capability of rotating about the boresight axis, called Deck ($DK$).
The primary science goal of BICEP3 is to search for signatures of cosmic inflation which would manifest as B-mode polarization in the Cosmic Microwave Background \cite{2014B3Zeesh}.
To date, no external calibration of the polarization axes of the BICEP3 detectors has been conducted.
While this has no effect on our ability to detect primordial B-modes, we can not distinguish the source of uniform polarization angle offsets measured through EB/TB minimization.
For this purpose, we have calibrated the absolute polarization properties of BICEP3 and the results reported in these proceedings will be used to provide our own birefringence constraints.

\section{FAR-FIELD POLARIZATION MEASUREMENTS}
\label{sec:ffpm}
\begin{figure}
    \centering

    \begin{adjustbox}{width={\textwidth},totalheight={\textheight},keepaspectratio}
    \includegraphics{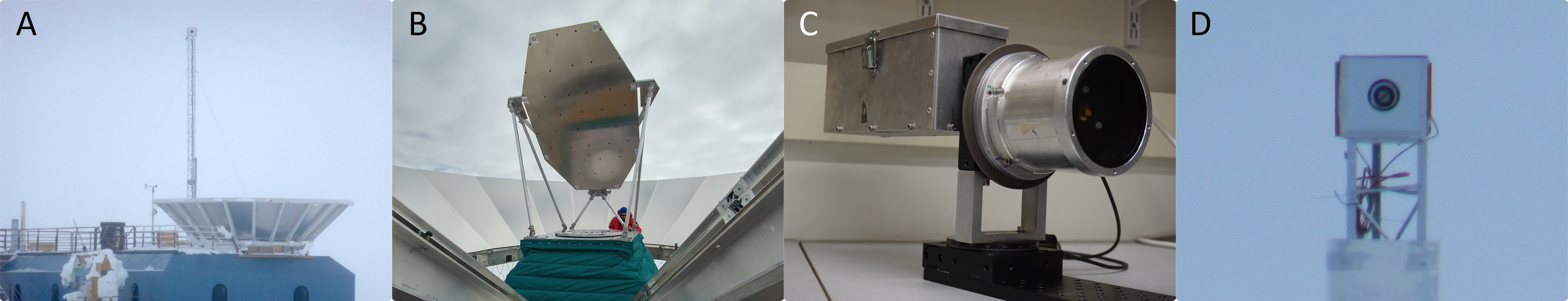}
    \end{adjustbox}
    \caption{Example images of the typical RPS calibration campaign. \textbf{A:} RPS deployed atop a calibration mast, secured by guy-lines. \textbf{B:} BICEP3 with calibration mirror installed. \textbf{C:} RPS benchtop setup - broadband noise source with absorptive shroud and wire grid secured to precision rotation stage. The tilt meter (not shown) is fixed to the same rigid aluminum plate as the rotation stage. \textbf{D:} Zoomed image of RPS installed in environmental enclosure shows red alignment strips which allows for a ($<1^\circ$) alignment with the telescope. Images courtesy of the BICEP Collaboration.}
    \label{fig:rpsimages}
\end{figure}

We map polarization response in the far-field using a Rotating Polarized Source (RPS) which is an electrically chopped, non-thermal, broad spectrum noise source (BSNS) fixed to a precision rotation stage and highly polarized using a wire grid (Fig. \ref{fig:rpsimages}).
The RPS is deployed atop a 12m mast on the Martin A. Pomerantz Observatory and observed by BICEP3 in the Dark Sector Laboratory 211m away\footnote{The far-field of BICEP3 is $\sim180$m}.
Even atop the mast, the RPS is well below BICEP3's normal elevation range so a flat, beam-filling, aluminum-honeycomb mirror is installed at a $45^\circ$ tilt to redirect rays onto the horizon when the telescope is pointed at zenith.
To accommodate the mirror, the co-moving absorbing forebaffle is removed.

The RPS is housed within an environmental enclosure and PID temperature controlled to ensure stability of the source power.
The environmental enclosure is equipped with high-visibility tabs on each side to allow for precise alignment ($<1^\circ$) with the telescope (see Figure \ref{fig:rpsimages}d).
We reference the polarization orientation of the source to zenith using a precision tilt meter installed alongside the source within the environmental enclosure.
The tilt of the source is zeroed and locked in using guy wires on the mast to which it is fixed.
Because the horizon is thermally bright, the RPS is electronically chopped at 20 Hz and demodulated during analysis to isolate the signal from the surrounding area.

\subsection{Observations}
\begin{figure}
    \centering

    \begin{adjustbox}{width={\textwidth},totalheight={\textheight},keepaspectratio}
    \includegraphics{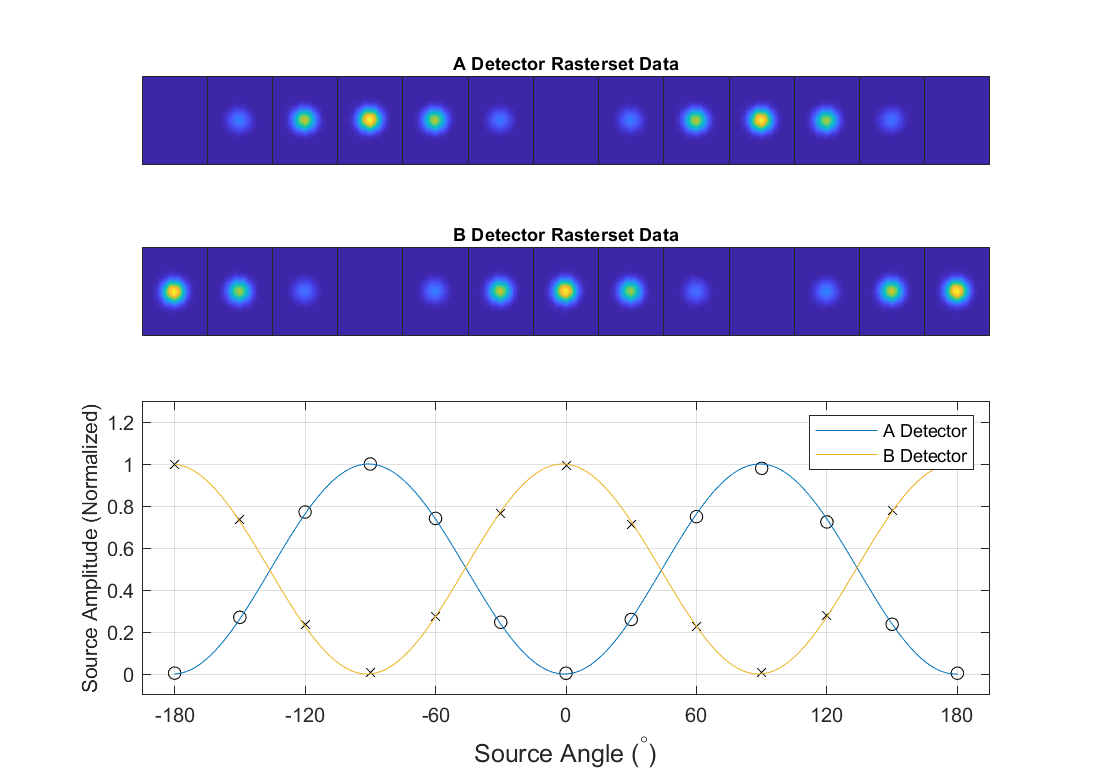}
    \end{adjustbox}
    \caption{Binned and smoothed beam maps of individual RPS rasters from a rasterset for an A-polarization detector (\textbf{Top}) and B-Polarization detector (\textbf{Middle}) and resulting modulation curve as a function of source command angle \textbf{(bottom)} where the {\color{NavyBlue}blue} and {\color{olive}yellow} lines are fits to the A- and B-polarization detectors respectively. Binning and smoothing is for visualization only as beams are fit in sample-space during analysis.}
    \label{fig:modcurve}
\end{figure}

With the RPS fixed at one polarization angle, we map polarized beams by rastering across the source $9^\circ$ in azimuth at $1.5^\circ/s$ and step up to $2^\circ$ in elevation in $0.1^\circ$ steps.
A full RPS observation, called a rasterset, consists of 13 separate rasters where the RPS is commanded from $-180^\circ$ to $180^\circ$ in $30^\circ$ increments.
This scan strategy allows us to sufficiently map the entire focal plane with multiple samples per beam over all source angles between cryogenic cycling.
An example of the modulation curve as a function of RPS angle is shown in Fig. \ref{fig:modcurve}.

\subsection{Dataset}
The RPS observations used in this analysis were taken in January and February of 2018.
Five observations were taken over four boresight angles of $1.25^\circ$, $46.25^\circ$, $91.25^\circ$, $136.25^\circ$ with an extra observation taken at $46.25^\circ$. The BICEP3 focal plane is clocked $1.25^\circ$ with respect to the $DK=0^\circ$ and so we add $1.25^\circ$ to ensure we closely observe maximum and minimum polarization coupling to the source. 
Observations over multiple boresight angles provides a consistency check of our measurement repeatability.

\section{POLARIZATION PARAMETERS}
\label{sec:analysis}
This section outlines the process through which we estimate polarization parameters.
To summarize, for each detector we fit a 2-D elliptical Gaussian profile across all rasters in a rasterset simultaneously.
The resulting amplitudes as a function of source angle are fit to a polarization response model.
Finally, the polarization parameters of the two orthogonal detectors corresponding to a single detector pair are combined into a per-pair polarization angle and cross-polarization response.

We fit a 2-D elliptical Gaussian profile across all beams in a rasterset simultaneously where the beam amplitude $A$ is a free parameter for each raster but the beam center $(x_0\,\,y_0)$\footnote{We use the same instrument-fixed coordinate system which is described in detail in our BICEP3 Beam Characterization Proceedings\cite{2016B3Karkare}.}, beam widths $(\sigma_x,\,\sigma_y)$, and correlation coefficient $\rho$ are single parameters fit across the entire rasterset.
\begin{equation}
    B_i(\mathbf{x}) = A_i e^{(\mathbf{x}-\mathbf{\mu})^T\Sigma^{-1}(\mathbf{x}-\mathbf{\mu})}
\end{equation}
Where $\mathbf{\mu} = (x_0\,\,y_0)$ is the beam center and 
\begin{equation}
\Sigma =
\begin{bmatrix} \sigma_x^2& \rho \sigma_x \sigma_y \\ \rho \sigma_x \sigma_y & \sigma_y^2 \end{bmatrix}
\end{equation}
This is to mitigate erroneous fits when the RPS and detector are $90^\circ$ out of phase and the constraining power on the beam is poor.

The array of amplitudes $A$ resulting from the Gaussian fits produces a modulation curve as a function of source angle $\theta$.
We fit a sinusoidal polarization response function to the modulation curves of the form
\begin{equation}
    A(\theta) = G\left(\cos\left(2\left(\theta+\psi\right)\right)-\frac{\epsilon+1}{\epsilon-1}\right)\left(N_1\cos(\theta)+N_2\sin(\theta)\right)
\end{equation}
where $G$ is the gain of the detector, $\psi$ is the detector polarization angle with respect to the source, $\epsilon$ is the detector cross-polarization response, and ($N_1,\,N_2$) are nuisance parameters describing a miscollimation of the source rotation axis.
We relate the detector polarization angle to the telescope zero $\phi_d$ by adding the polarization orientation of the source with respect to the telescope $\phi_s$ which is computed through a pointing model.
\begin{equation}
    \phi_d \equiv \psi + \phi_s
\end{equation}

Finally, we compute the polarization parameters of a detector pair.
For a pair of detectors A and B, the response in Stokes Q and U of the detector pair is calculated as
\begin{equation}\label{eq:QU}
\begin{split}
    &Q = \frac{\left[\cos(2\phi_A)-\epsilon_A\cos(2\phi_A)\right]-\left[\cos(2\phi_B)-\epsilon_B\cos(2\phi_B)\right]}{2+\epsilon_A+\epsilon_B}\\
    &\\
    &U = \frac{\left[\sin(2\phi_A)-\epsilon_A\sin(2\phi_A)\right]-\left[\sin(2\phi_B)-\epsilon_B\sin(2\phi_B)\right]}{2+\epsilon_A+\epsilon_B}\\
\end{split}
\end{equation}
and the effective per-pair polarization angle $\phi_{pair}$ and cross-polarization response $\epsilon_{pair}$ is simply
\begin{equation}\label{eq:phiq}
\begin{split}
    &\phi_{pair} = \frac{1}{2}\tan^{-1}\frac{U}{Q}\\
    &\\
    &\epsilon_{pair} = 1-\sqrt{Q^2+U^2}
\end{split}
\end{equation}

\subsection{Results and Discussion}\label{sub:results}
\begin{figure}[!h]
    \centering
    \includegraphics[width=1\linewidth]{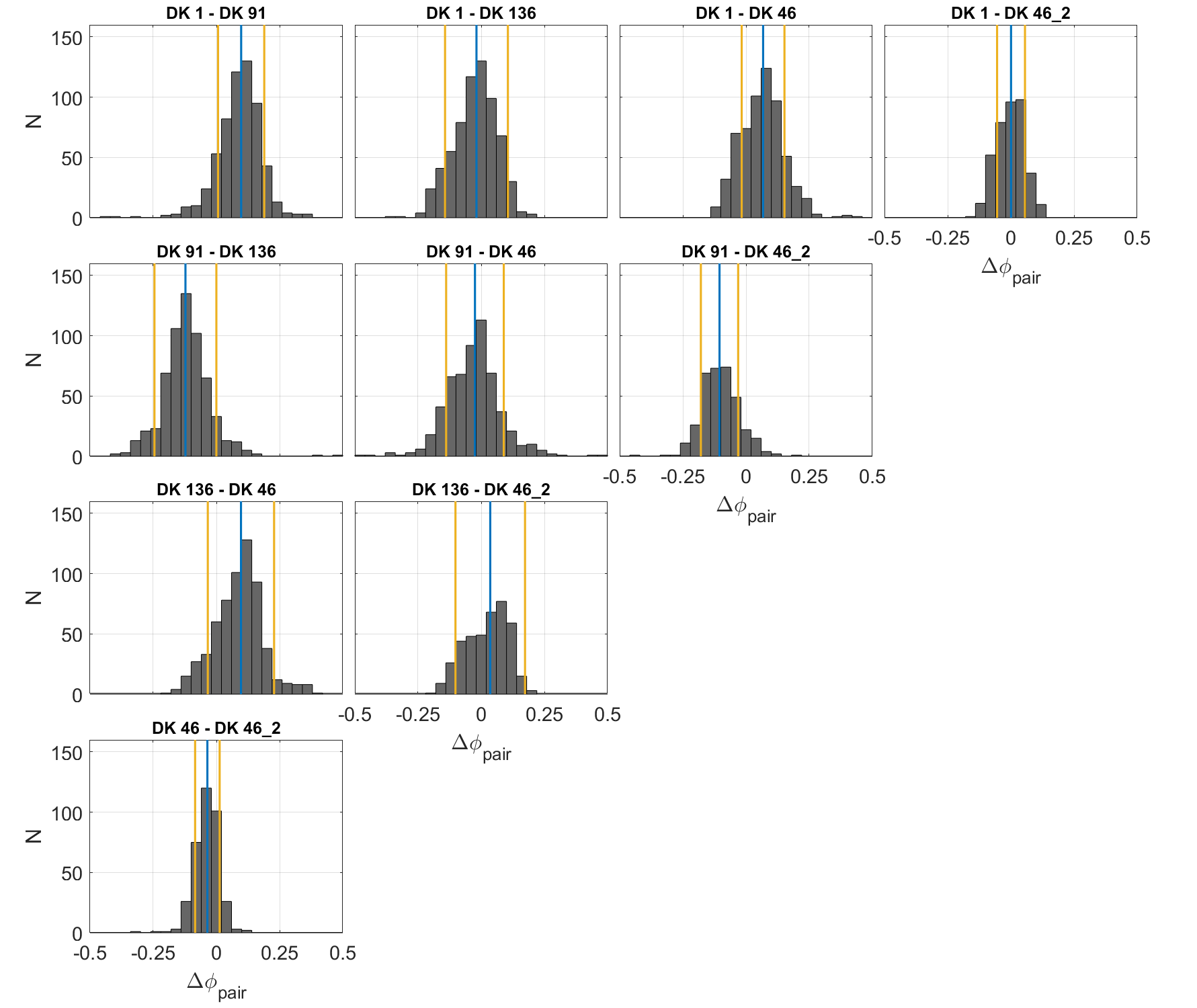}

    
    \caption{Histograms of \phip{} differenced between boresight angles where DK 46 and DK 46\_2 distinguish two different datasets at the same boresight angle. The {\color{NavyBlue}blue} lines indicate the median. The {\color{olive}yellow} lines indicate the 1$\sigma$ standard deviations. We report the STD/$\sqrt{2}$ of the DK46-DK46\_2 distribution as the characteristic statistical uncertainty of $\pm0.035^\circ$. We report the systematic uncertainty of the absolute polarization angle as the standard deviation on the medians which is $\pm0.075^\circ$.}
    \label{fig:consist}
\end{figure}

\begin{figure}
    \centering
    \includegraphics[width=1\linewidth]{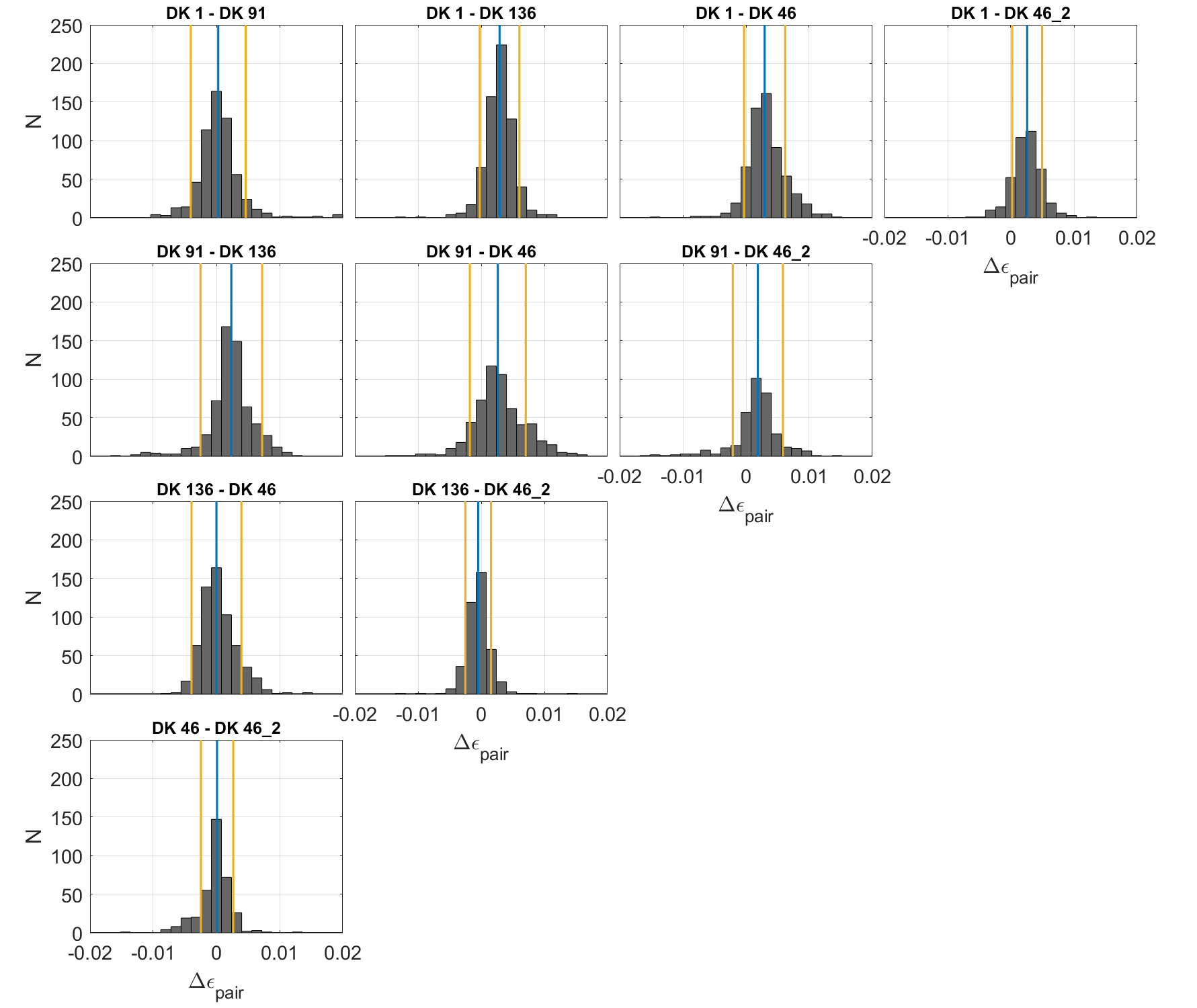}
    
    \caption{Histograms of \epsp{} differenced between boresight angles where DK 46 and DK 46\_2 distinguish two different datasets at the same boresight angle. The {\color{NavyBlue}blue} lines indicate the median. The {\color{olive}yellow} lines indicate the 1$\sigma$ standard deviations. We report the characteristic statistical and systematics uncertainties using the same methods for \phip{} (Fig. \ref{fig:consist}) of $\pm0.0014$ and $\pm0.0018$ respectively.}
    \label{fig:consist2}
\end{figure}

For the results presented in these proceedings, we estimate \phip{} and \epsp{} at each boresight angle before averaging across all boresight angles.
Of the 1200 possible detector pairs, 827 of those are considered science-grade and we have characterized properties for 765 of those pairs.
We find some systematic fluctuation of the uniform polarization offset between all datasets as evidenced in Fig. \ref{fig:consist} which we ascribe as the result from uncertainties in the mirror orientation in the pointing model.
As such, we compute the median of the differences between parameter estimates of the five datasets and report the $1\sigma$ deviations on the medians to be $0.075^\circ$ and $0.0018$ as the systematic uncertainty for \phip{} and \epsp{} respectively (see Fig. \ref{fig:consist} and Fig. \ref{fig:consist2} respectively).
In addition, we calculate the expected systematic uncertainty on the absolute polarization angle via error propagation through the pointing model and find it to be $\mathcal{O}(1^\circ)$.
Due to this inconsistency, we currently consider the uncertainty of $0.075^\circ$ to be a lower limit.
We report the per-pair $1\sigma$ statistical uncertainty of $\sigma_{\phi}=0.035^\circ$ and $\sigma_{\epsilon}=0.0014$ calculated as the STD/$\sqrt{2}$ of the difference between the two sets of parameter estimates made at the same $46.25^\circ$ boresight angle.
A summary of statistics for our results are shown in Table \ref{tab:results}.

\begin{figure}
    \begin{subfigure}{.49\textwidth}
    \centering
    \includegraphics[width=1\linewidth]{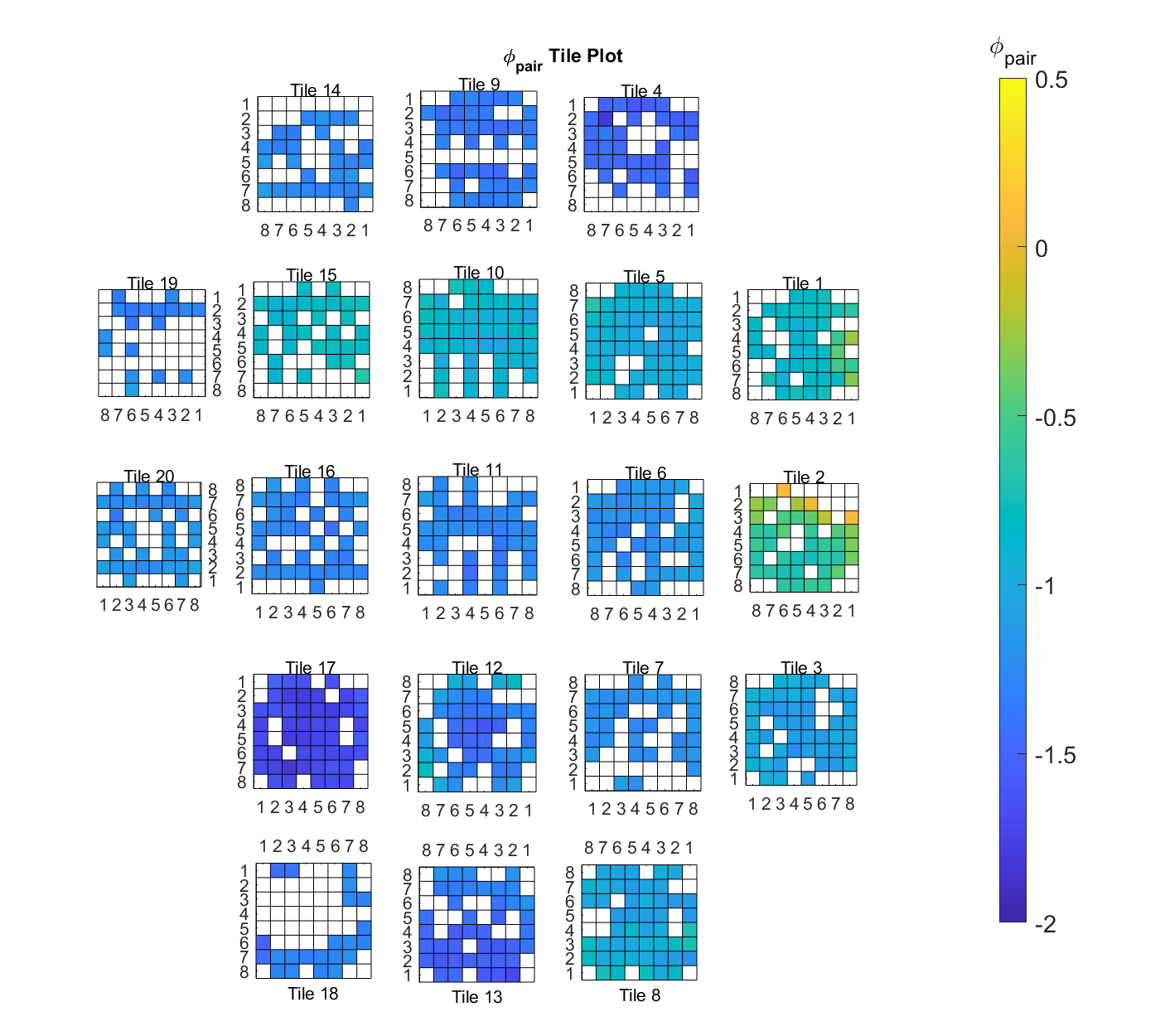}
    \end{subfigure}
    \begin{subfigure}{.49\textwidth}
    \centering
    \includegraphics[width=1\linewidth]{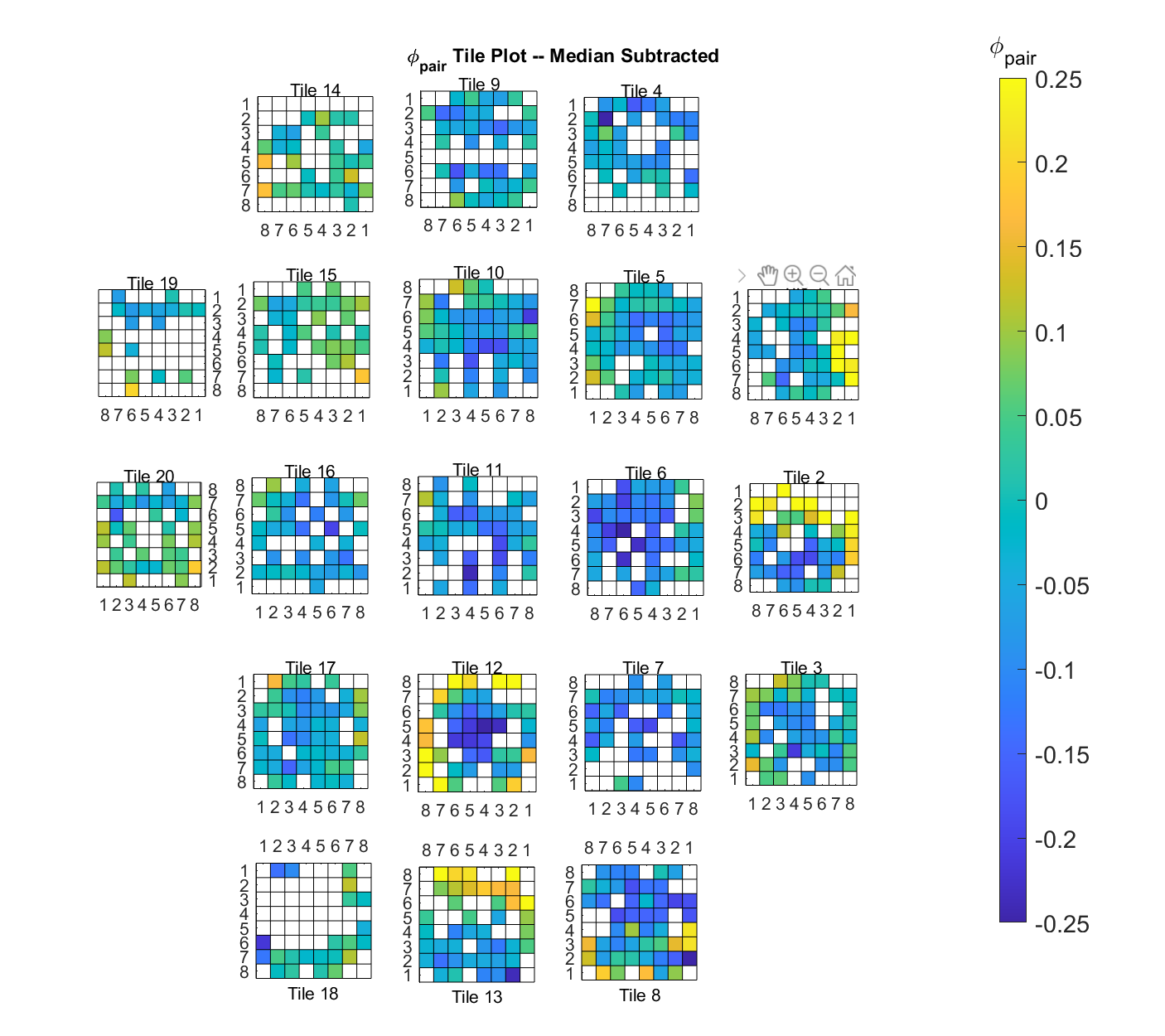}
    \end{subfigure}
    
    
    \begin{subfigure}{.49\textwidth}
    \centering
    \includegraphics[width=1\linewidth]{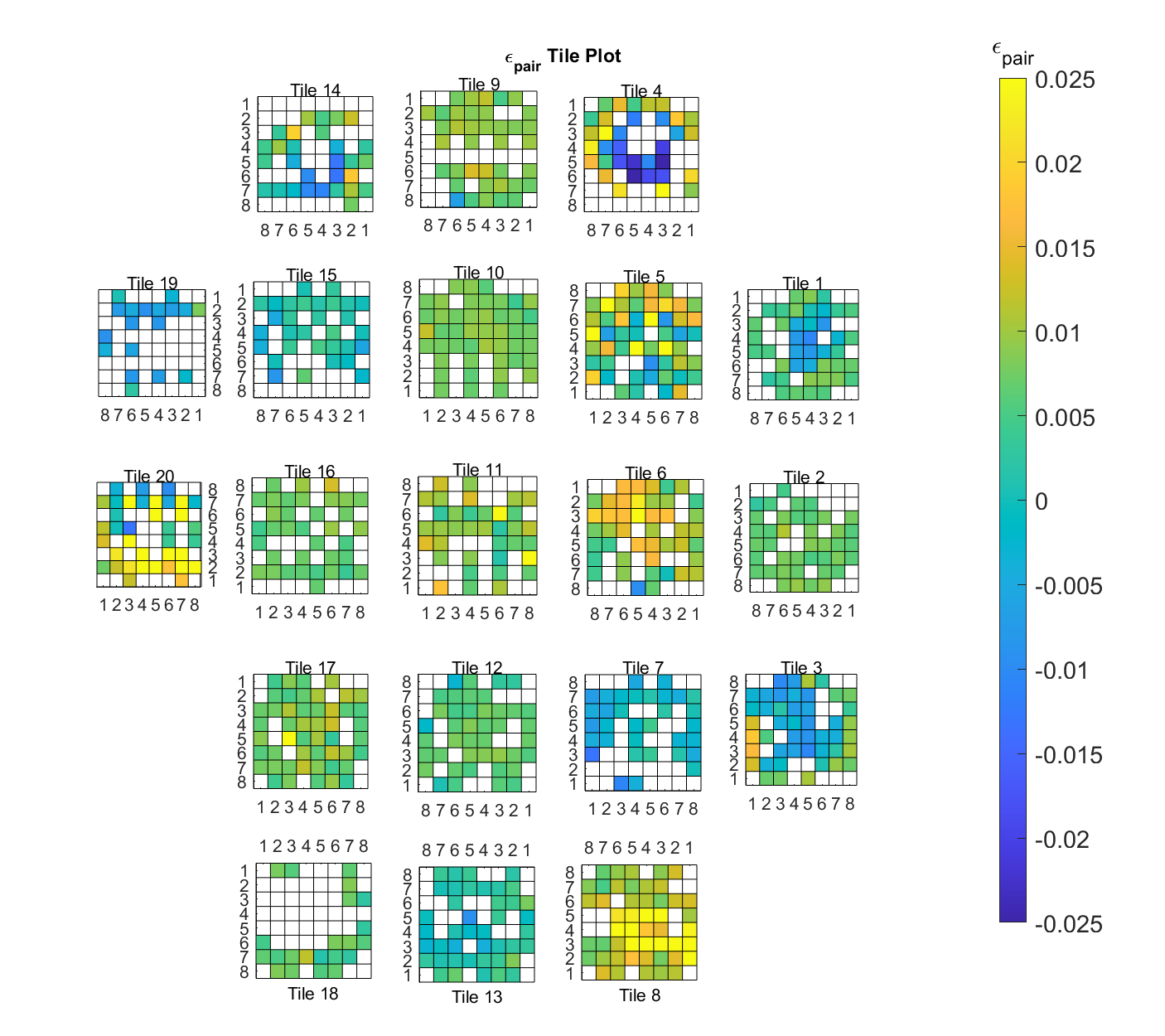}
    \end{subfigure}
    \begin{subfigure}{.49\textwidth}
    \centering
    \includegraphics[width=1\linewidth]{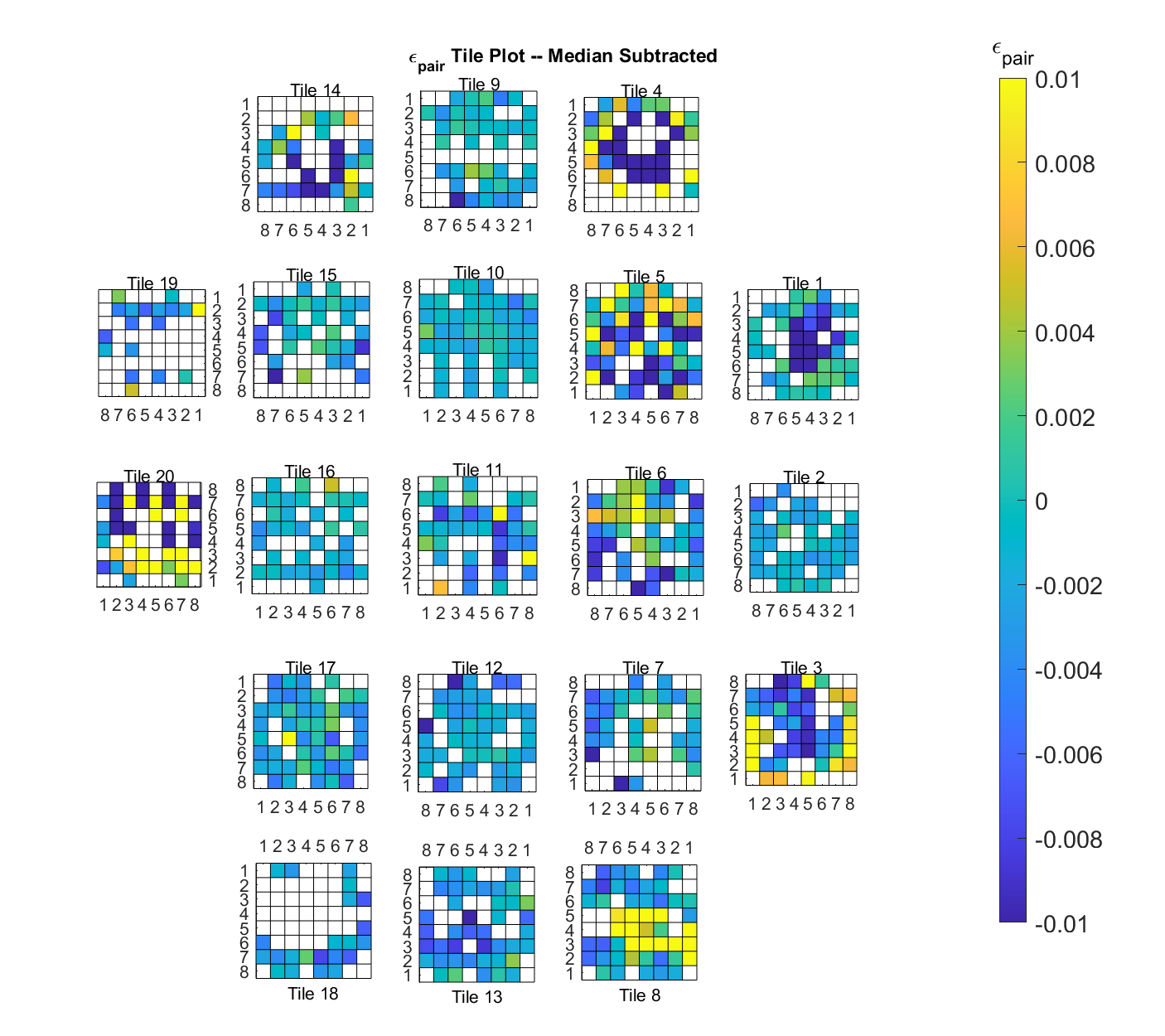}
    \end{subfigure}
    
    \caption{$\phi_{pair}$ (\textbf{Top Row}) and $\epsilon_{pair}$ (\textbf{Bottom Row}) across the focal plane. \textbf{Left Columns} and \textbf{Right Columns} are without and with the median subtracted from each tile respectively.}
    \label{fig:results}
\end{figure}

We show results for \phip{} and \epsp{} across the focal plane in Fig. \ref{fig:results}.
We find coherent offsets in \phip{} across individual tiles with a tile-to-tile scatter of $0.32^\circ$ which is large compared to an overall FPU scatter of $0.13^\circ$ when the median is subtracted from each tile.
At the individual tile level, most detectors appear to be distributed randomly -- the notable exceptions being Tiles 1, 2, 12, and 13 which exhibit large deviations at one or all tile edges.
We note that while we report an overall polarization angle of $-1.20^\circ$, this should not be considered the absolute polarization calibration of the instrument.
The actual absolute uniform polarization angle will be the mean contribution from the detector pairs weighted by their individual sensitivities which we do not consider here.

We find cross-polarization response in the majority of detector pairs to be \epsp$<0.9\%$ with $\sim20\%$ of pairs showing negative cross-polarization response.
Detector pairs with negative cross-polarization response appear to be localized to certain tiles which indicates that these values could be the result of poor fits driven by systematics intrinsic to those tiles.
While most tiles appear to be randomly distributed, we also find structure in \epsp{} within specific tiles (1, 3, 4, and 8 for instance), though this structure does not appear to be correlated with the structure observed in \phip.

\begin{table}
    \centering
    \begin{adjustbox}{width={\textwidth},totalheight={\textheight},keepaspectratio}
    \begin{tabular}{cccccc}
    \hline
    \hline
    \textbf{Parameter} & \textbf{FPU Median Per-Tile} & \textbf{FPU Scatter Per-Tile} & \textbf{FPU Scatter Per-Pair} & \textbf{Stat. Uncert.} & \textbf{Sys. Uncert.}\\
    \hline
    \hline
    \phip{}$\,\left(^\circ\right)$ &-1.20&0.32&0.13&0.035&0.075\\
    \hline
    \epsp{} &0.0069&0.0097&0.0084&0.0014&0.0018\\
    \hline
    \hline
    \end{tabular}
    \end{adjustbox}
    \caption{Summary of statistics for \phip{} and \epsp{}. \textbf{FPU Median Per-Tile:} Median across the whole focal plane based on per-tile medians. \textbf{FPU Scatter Per-Tile:} Standard deviation across the focal plane based on per-tile medians. \textbf{FPU Scatter per-pair:} Standard deviation across the focal plane based on individual detector pairs with per-tile medians subtracted. \textbf{Stat. Uncert. / Sys. Uncert.:} Characteristic statistical and systematic uncertainty per-pair, see \S \ref{sub:results} for details.}
    
    \label{tab:results}
\end{table}

\section{CONCLUSIONS}
\label{sec:conc}
In these proceedings we have presented high signal-to-noise polarization calibration data of BICEP3's $\sim800$ detector pairs from January 2018.
We report measurements of polarization angles which shows coherent scatter between focal plane tiles with a comparatively smaller (factor of 2) pair-to-pair variation within each given tile.
While the properties of detector pairs appear to be randomly distributed we do find structure in some tiles to high-significance possibly caused by impedance mismatch at the tile edges or driven by systematics in detector performance intrinsic to a given tile.

While our results demonstrate the high precision with which we can measure polarization properties, the relatively large systematic uncertainty on the absolute calibration calculated from error propagation requires further understanding.
In future work, our plan is to focus on reducing the uncertainties in the pointing model which we expect will improve the systematic uncertainty on the absolute polarization angle.
In combination with BICEP3 CMB data, we can then use these results to provide competitive constraints on cosmic birefringence.

%

\section{ACKNOWLEDGEMENTS}
The BICEP/\textit{Keck} project (including BICEP2, BICEP3 and BICEP Array) have been made possible through a series of grants from the National Science Foundation including 0742818, 0742592, 1044978, 1110087, 1145172, 1145143, 1145248, 1639040, 1638957, 1638978, 1638970, 1726917, 1313010, 1313062, 1313158, 1313287, 0960243, 1836010, 1056465, \& 1255358 and by the Keck Foundation. The development of antenna-coupled detector technology was supported by the JPL Research and Technology Development Fund and NASA Grants 06-ARPA206-0040, 10-SAT10-0017, 12-SAT12-0031, 14-SAT14-0009, 16-SAT16-0002, \& 18-SAT18-0017. The development and testing of focal planes were supported by the Gordon and Betty Moore Foundation at Caltech. Readout electronics were supported by a Canada Foundation for Innovation grant to UBC. The computations in this paper were run on the Odyssey cluster supported by the FAS Science Division Research Computing Group at Harvard University. The analysis effort at Stanford and SLAC was partially supported by the Department of Energy, Contract DE-AC02-76SF00515. We thank the staff of the U.S. Antarctic Program and in particular the South Pole Station without whose help this research would not have been possible. Tireless administrative support was provided by Kathy Deniston, Sheri Stoll, Irene Coyle, Donna Hernandez, and Dana Volponi.

\bibliography{report} 
\bibliographystyle{spiebib} 

\end{document}